\documentclass[conference]{IEEEtran}

\usepackage{cite}          % [1], [2]-[5] compression
\usepackage{amsmath,amssymb}
\usepackage{graphicx}
\graphicspath{{figures/}}
\usepackage{booktabs}      % tables
\usepackage[hidelinks]{hyperref}

\newcommand{\state}{\mathbf{x}}
\newcommand{\params}{\boldsymbol{\theta}}
\newcommand{\coord}{\boldsymbol{\eta}}          % normalised coordinates of (theta, x0)
\newcommand{\est}[1]{\hat{#1}}
\newcommand{\expect}{\mathbb{E}}
\DeclareMathOperator*{\argmin}{arg\,min}

\usepackage{xcolor}
\begin{document}

\title{A risk-aware individualized blood pressure target to avoid hypotension in surgery}

\author{%
\IEEEauthorblockN{Simon Buré\IEEEauthorrefmark{1}, Jona Joachim\IEEEauthorrefmark{2}\IEEEauthorrefmark{3}, Florent Di Meglio\IEEEauthorrefmark{1}}
\IEEEauthorblockA{\IEEEauthorrefmark{1}Centre Automatique et Systèmes, Mines Paris -- PSL, Paris, France, \IEEEauthorrefmark{2}Université Paris Cité,\\ \IEEEauthorrefmark{3}Service of Biochemistry and Molecular
Biology, St-Louis-Lariboisière-Fernand Widal
University Hospitals, Paris, France \\
\texttt{simon.bure@minesparis.psl.eu}}
}

\maketitle
\thispagestyle{plain}
\pagestyle{plain}

\begin{abstract}
We propose an online method for choosing both a blood pressure target and a
norepinephrine dose for an individual surgical patient. From a short record
of mean arterial pressure (MAP) and norepinephrine delivery, the method fits
a reduced pharmacokinetic/pharmacodynamic model and uses a local Gaussian
approximation to propagate uncertainty to future MAP readings. It searches
for the lowest target for which a bolus and constant infusion keep the
predicted probability of at least one MAP reading below 65~mmHg after the
rise phase within a prescribed limit. If no such target is found, the
decision returns to the clinician. We assess calibration of the
predicted minimum MAP on retrospective surgical records under the doses
actually administered. In a cohort of 4{,}096 virtual patients, the method
recommended a dose for 74\,\% of patients at a nominal risk of 5\,\%.
Among the patients receiving a recommendation, 6.9\,\% had a reading below
65~mmHg. A fixed target of 70~mmHg yielded a rate of 25\,\% when applied to
the full virtual cohort.
\end{abstract}

%% ~1.25 columns-pair (approx. 1 page of 6). Budget: 4 paragraphs.
\section{Introduction}
\label{sec:intro}

This paper proposes an online protocol that individualises blood pressure management for surgery patients. We propose to calibrate a response model to a vasopressor,
NorepinEphrine (NE), from a few minutes of Mean Arterial Pressure (MAP)
readings. Subsequently, we propose a NE dose that reaches the lowest MAP target with a risk of hypotension for the patient lower than a chosen error-level $\alpha$.
Individualisation from data comes with uncertainty, that we compute, and propagate to the dosing decision.
The uncertainty calibration is checked on real data from $31$ patients undergoing surgery, and the full protocol is evaluated \textit{in silico} against an oracle and a non-individualised target on $4{,}096$ virtual patients.

% --- P1: problem ----------------------------------------------------------
Hypotension is frequent during general anaesthesia, and even short episodes
with a MAP below 55 to 65~mmHg are associated with
acute kidney injury, myocardial injury and
death~\cite{walsh_relationship_2013,joosten_intraoperative_2026}.
International recommendations therefore ask to keep the MAP above 60 to
65~mmHg~\cite{joosten_intraoperative_2026}, and targets set for each patient
improve postoperative outcomes~\cite{futier_effect_2017}.
NE, given as boluses (large doses over short periods of time) or as a continuous infusion, is the
vasopressor most used for this, and its dose is titrated by hand. Patients
respond very differently to the same dose: after a bolus, the MAP of the
patients of~\cite{joachimDesignPharmacokineticPharmacodynamic2024} rises by
24\,\% in median, and by 15 to 31\,\% for the middle half. A dose that suits
the typical patient can thus leave one patient hypotensive and push another
far above the target.

% --- P2: literature -------------------------------------------------------
Pharmacokinetic/pharmacodynamic (PK/PD) models typically describe this response.
Models of NE exist for septic shock and trauma
patients~\cite{beloeil_norepinephrine_2005} and for healthy volunteers before
and during anaesthesia~\cite{li_population_2024}. In~\cite{joachimDesignPharmacokineticPharmacodynamic2024}, a population model is built for low-dose peripheral NE during surgery, from plasma
concentrations (measured from blood samples after the surgery is over) and MAP (measured during surgery from an invasive sensor). The population model can be used as a dsitribution prior, but poorly predicts a given individual patient.

Regarding blood pressure management, closed-loop vasopressor systems
automatically adjust NE infusion rates using measured
MAP~\cite{merouani_norepinephrine_2008,joosten_automated_2019},
reducing the time spent in hypotension in randomised
trials~\cite{joosten_automated_2021,desebbe_control_2022}.
{Such controllers can accommodate interpatient variability through
robust control~\cite{rinehart_closed-loop_2019} or adapt to it online, with a
neural network predictor in animals~\cite{kashihara_adaptive_2004} or a
data-driven digital twin of critically ill
patients~\cite{nair_artificial_2026}. In all of them the target is fixed, and
the uncertainty of the patient's model does not enter the dose.}
Another approach is to individualise a model from the patient's own
response, a longstanding idea in
pharmacokinetics~\cite{sheinerBayesianIndividualizationPharmacokinetics1982}.
Online identification has been explored for
phenylephrine~\cite{luspay_adaptive_2016} and for propofol and
remifentanil~\cite{da_silva_online_2010}, including in combination
with model predictive control~\cite{aubouinpairault_online_2024}.
These identification studies concern drugs other than NE, and standard
pharmacological model structures can be difficult to identify from
early clinical data~\cite{wahlquist_identifiability_2021}.
This motivates a complementary approach to NE dosing that
individualises a model from MAP measurements and provides dose
recommendations for clinician approval.

% --- P3: gap --------------------------------------------------------------
To this end, two ingredients are missing. First, the approach of~\cite{joachimDesignPharmacokineticPharmacodynamic2024} relies on plasma
concentrations, which cannot be instantly measured and are unvailable during surgery. Second, a model fitted
on a few minutes of MAP readings is uncertain, to a degree that depends on the
patient and on the data, yet most dosing schemes either ignore this
uncertainty when the dose is chosen or cover it with a margin designed for the
whole population. We propose (i) to calibrate an individual model estimable from the
MAP alone, and (ii) a dosing rule whose safety margin follows the uncertainty of
the patient's own fit.

% --- P4: contributions + roadmap -------------------------------------------
{The main contribution of this paper is a dose recommendation that can
abstain: from a few minutes of MAP, it proposes the lowest target whose NE doses
keep the predicted risk of hypotension below $\alpha$, a bolus and a rate, or
returns the decision to the clinician when no target is safe.} Our approach relies on the following ingredients
\begin{itemize}
  \item a reduced NE PK/PD model whose seven parameters are estimated from MAP
        readings alone, at little cost in fit quality on the cohort
        of~\cite{joachimDesignPharmacokineticPharmacodynamic2024};
  \item from the fit's linearised posterior, the law of the coming MAP
        readings, and from it a personalised target: the lowest MAP target
        whose control input keeps the risk of a reading below 65~mmHg at most
        $\alpha$, or no injection when no target is safe;
  \item a control input built from a bolus and an infusion, the two primitives of
        routine practice, by inverting the model;
\end{itemize}
We illustrate the merits of this approach on 155 forecast windows of the clinical cohort, and on 4{,}096 virtual patients against an oracle that knows the true patient.

The paper is organized as follows. Section~\ref{sec:protocol} presents the model, the target and the control input.
Section~\ref{sec:application} checks them on the clinical cohort, then on the
virtual one, and Section~\ref{sec:discussion} discusses the results and their
limits.

\section{Online Individualisation and Control Protocol}
\label{sec:protocol}

This section introduces the building blocks that we rely on for safe online identification and control of the MAP under NE.
We start by introducing the mathematical model of a patient, and our proposed modification in order to make its identification in routine surgery possible \& cheap (Section \ref{subsec:model}).
Then, we derive a first-order approximation of the distribution of future MAP values after calibration, and from it a personalised target. The latter is defined as the lowest MAP target for which the corresponding NE dose is considered safe at a chosen risk of hypotension ~(\ref{subsec:gauge}).  When no target is deemed safe, no reccomendation is made to the clinician.
Finally we propose an open-loop control law compatible with clinical practices that reaches a giving target (\ref{subsec:command}).

\subsection{Model simplification and calibration}
\label{subsec:model}

Consider the model from the NE PK/PD model of~\cite{joachimDesignPharmacokineticPharmacodynamic2024}. The state
$\state = (A_d, A_c, A_p)$ represents the exogenous NE amounts (nmol) in a depot, a
central and a peripheral compartment, driven by the drug infusion rate $u$ (nmol/s)
delayed by $T_{\mathrm{lag}}$:
\begin{equation}
  \label{eq:pk}
  \begin{aligned}
    \dot A_d &= u(t - T_{\mathrm{lag}}) - k_a A_d^{\beta}, \\
    \dot A_c &= k_a A_d^{\beta} - (k_{12} + k_{el}) A_c + k_{21} A_p, \\
    \dot A_p &= k_{12} A_c - k_{21} A_p .
  \end{aligned}
\end{equation}

The MAP (mmHg) depends on the total concentration
$C = C_{\mathrm{endo}} + A_c/V_c$, as follows:
\begin{equation}
  \label{eq:emax}
  y = E_0 + (E_{\max} - E_0)\,\frac{C^{\gamma}}{C^{\gamma} + EC_{50}^{\gamma}} .
\end{equation}
This complete model has twelve parameters. We refer the reader to~\cite{joachimDesignPharmacokineticPharmacodynamic2024} for their physical interpretation. To identify the parameters of ~\eqref{eq:pk},\eqref{eq:emax}, NE concentration and MAP measurements are gathered from 29 surgery patients.

\subsubsection{Model simplification}
Identifying the same model online poses two problems. First, concentration measurements are not available in routine surgery. This makes certain combinations of parameters indistinguishable. Second, identification must be as cheap as possible. Thus, we consider here a delay-free, linear version of \eqref{eq:pk}, i.e. we impose~$T_{\mathrm{lag}} = 0$ and $\beta = 1 $.  Further,~\ref{eq:emax} is replaced with a simplified function, whose parameters are estimable from MAP measurements as follows
\begin{equation}
  \label{eq:pd}
  y = h(A_c) = E_{\max} - \frac{a}{b + A_c} ,
\end{equation}

\begin{figure}[htb]
  \centering
  \includegraphics[width=\columnwidth]{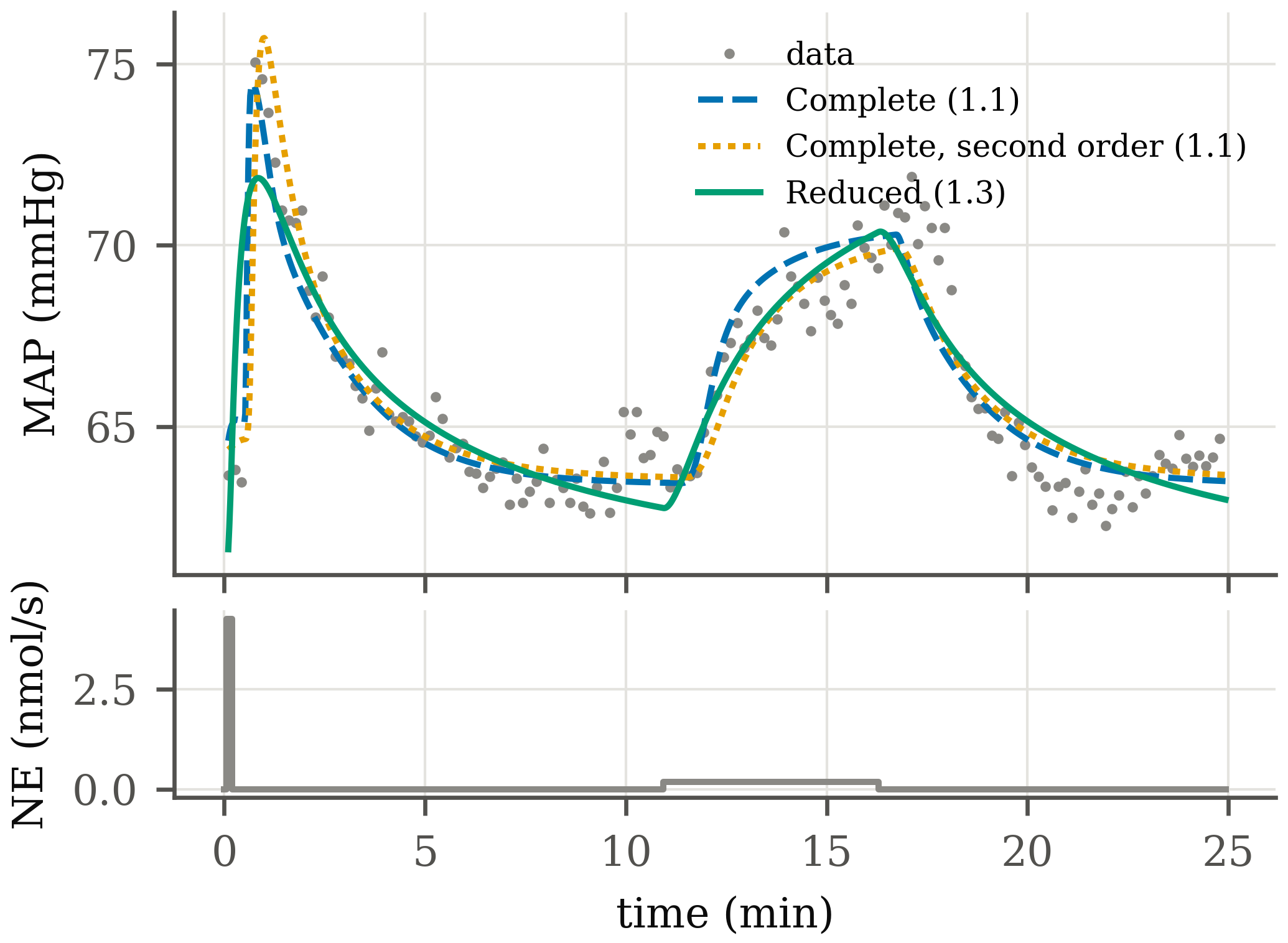}
  \caption{Whole-record fits of one patient of Joachim et al.'s cohort (least
    squares in the box $\mathcal{B}$ of~\eqref{eq:fit}, started from the population): MAP
    and NE infusion. In brackets, the patient's MAP RMSE (mmHg).}
  \label{fig:fits}
\end{figure}

To evaluate the potential for further reduction, we apply standard model ranking \& selection analysis relying on the different methods proposed in \cite{chu_improving_2009, wu_selection_2011, mclean_mathematical_2012, mclean_mean-squared-error_2012}. Data-based criteria of Wu et al. and McLean et al. keep all seven parameters of the reduced model \ref{eq:pk} \& \ref{eq:pd} while the criterion of Chu et al. fixes $k_{12}$ alone. We keep the seven parameters in the rest of the paper.
Figure~\ref{fig:fits} illustrates the performance of the simplified model by depicting the three models fitted to one patient of Joachim et al.'s cohort. Over the 32 patients, the MAP RMSE (mean $\pm$ standard deviation) is $2.33 \pm 1.50$~mmHg for the complete model~\eqref{eq:emax} (12 parameters, fitted to MAP and Cc), $2.24 \pm 0.88$ for its second-order variant (12 parameters) and $2.43 \pm 0.93$ for the reduced model~\eqref{eq:pd} (7 parameters, fitted to the MAP alone);
the Cc RMSE of the complete models is $1.76 \pm 1.47$ and $1.50 \pm 0.83$~nmol/L.

In this context, a patient is described by their seven parameters $\params$ and by their initial state
$\state_0 = \state(t_0)$, which is estimated as well to take into account
potential pre-recording injections. Both are estimated in normalised
coordinates $\coord \in \mathbb{R}^d$, with $d = 10$ for the reduced model,
centred on the population model
of~\cite{joachimDesignPharmacokineticPharmacodynamic2024} and scaled by its
inter-individual variability: the population model is $\coord = 0$, and
$\mathcal{N}(0, I)$ is the prior of $\coord$ (Appendix~\ref{app:coord}).
\subsubsection{Model identification}
At a given time $t$, let $y_i$ be the $n_t$ MAP samples at times
$t_i \le t$, and $\hat y(t_i; \coord)$ the output of~\eqref{eq:pk}--\eqref{eq:pd}
under the recorded infusion. The individual estimate is the maximum a
posteriori one~\cite{sheinerBayesianIndividualizationPharmacokinetics1982},
\begin{equation}
  \label{eq:fit}
  \begin{aligned}
  \est{\coord}_t &= \argmin_{\coord \in \mathcal{B}} J_t(\coord), \\
  J_t(\coord) &= \frac{1}{\sigma_t^2} \sum_{t_i \le t} \bigl(y_i - \hat y(t_i; \coord)\bigr)^2
  + \lVert \coord \rVert^2 ,
  \end{aligned}
\end{equation}
where the box $\mathcal{B}$ keeps the solution in a biologically reasonable
region (Appendix~\ref{app:coord}).

From this point on, we write $\est{\params}_t$ for the parameters of $\est{\coord}_t$ and $\est{\state}(t)$
for the state they give at $t$.

\subsection{Forecast Uncertainty and a Safe Personalised Target}
\label{subsec:gauge}
The individual point estimate $\est{\coord}_t$ gives a forecast of the MAP under an injection $u$ and for a given time window $W_t$.
This forecast is then used to recommend a safe injection, but it comes with uncertainty: the model has only seen a few minutes of data, is only an approximation of a patient, and readings are noisy.
Uncertainty is patient-specific, and computing it is of paramount importance to ensure patients do not get recommended doses that could lead to hypotension.
In the following, we derive a first-order approximation of the probability distribution of the future MAP readings from $\est{\coord}_t$, and, from this, the personalised target which we define as the lowest MAP target whose injection avoids hypotension with a given risk $\alpha$.

The patient's coordinates $\coord$ are unknown, and their posterior
distribution describes the uncertainty remaining after observing the
readings up to $t$.
Treating the noise level $\sigma_t$ as fixed, the Gaussian prior
$\mathcal{N}(0,I)$ restricted to $\mathcal{B}$ and the independent
Gaussian observation noise assumed in~\eqref{eq:fit} give
\[
p(\coord \mid y_1,\dots,y_{n_t},\sigma_t)
\propto \exp\bigl(-J_t(\coord)/2\bigr),
\qquad \coord \in \mathcal{B}.
\]
The estimate $\est{\coord}_t$ is a mode of this posterior density.
Defining the estimation error as
$\delta\coord = \coord - \est{\coord}_t$, its conditional distribution
is obtained by translating the posterior by $-\est{\coord}_t$.
This distribution quantifies the uncertainty in the estimated
patient coordinates. The posterior has no closed form, because the model output $\hat y$ is
nonlinear in $\coord$. A first-order expansion of the output around the
estimate as follows
\begin{equation}
  \label{eq:lin}
  \hat y(\tau; \est{\coord}_t + \delta\coord) \approx \hat y(\tau; \est{\coord}_t)
  + s_t(\tau)^\top \delta\coord ,
\end{equation}
where $s_t(\tau) = \partial \hat y(\tau; \coord)/\partial \coord$ at
$\est{\coord}_t$ is obtained by automatic
differentiation~\cite{anderssonCasADiSoftwareFramework2019} of the exact
solution of the linear~\eqref{eq:pk} (with $\beta = 1$ and $T_{\mathrm{lag}} = 0$,
it admits a closed form solution under piecewise-constant infusion).
~\eqref{eq:lin} makes $J_t$ quadratic around the estimate:
\begin{equation}
  \label{eq:post}
  \begin{aligned}
  J_t(\est{\coord}_t + \delta\coord) &\approx J_t(\est{\coord}_t)
  + \delta\coord^\top P_t^{-1}\, \delta\coord , \\
  P_t &= \bigl(S_t^\top S_t/\sigma_t^2 + I\bigr)^{-1},
  \end{aligned}
\end{equation}
where $S_t \in \mathbb{R}^{n_t \times d}$ stacks the rows $s_t(t_i)^\top$ of the
readings up to $t$. The posterior is therefore approximately
Gaussian~\cite{mackay_bayesian_1992},
$\delta\coord \stackrel{\text{approx.}}{\sim} \mathcal{N}(0, P_t)$. This expansion is the only approximation of the section.
Two simplifications come on top of it. The true posterior is truncated by the
bounds of $\mathcal{B}$, which the Gaussian ignores. And $P_t$ uses the noise
level $\sigma_t$ as if it were known, although it is estimated from the same
readings.

The noise variance $\sigma_t^2$ is estimated from the same readings, by
maximising their marginal likelihood under the
expansion~\cite{mackay_bayesian_1992} (Appendix~\ref{app:sigma}).

The same expansion turns the uncertainty of the fit into a forecast. Consider
an infusion $u$, and the window $W_t = (t + T_r, t + 10\,\mathrm{min}]$ that
follows the rise of a control input (Section~\ref{subsec:command}). Its $n'$
readings $Y_j$, at times $t_j$, are approximately
\begin{equation}
  \label{eq:readings}
  Y_j \approx \hat y_t(t_j; u) + [S'_t\, \delta\coord]_j + \varepsilon_j ,
\end{equation}
where $S'_t$ stacks the rows $s_t(t_j)^\top$ under $u$, and
$\varepsilon_j \sim \mathcal{N}(0, \sigma_t^2)$ is the measurement noise, white
and independent of $\delta\coord$. The readings thus follow, approximately, the
Gaussian law $\mathcal{N}(\hat y_t, \Sigma_t)$, with
$\Sigma_t = S'_t P_t S_t'^{\top} + \sigma_t^2 I$: the uncertainty of the fit,
shared by every reading of the window, plus the noise, independent from one
reading to the next. We call it the forecast law. It ignores the error of the
reduced model itself and any correlation of the noise, so whether real readings
follow it is a question for the data (Section~\ref{subsec:real}).

The forecast law is used in two ways, depending on whether we deal with previously collected data or a new patient. The trace of the forecast covariance gives its expected squared error
over the window, that is to say the gauge
\begin{equation}
  \label{eq:gauge}
  V_t = \frac{1}{n'} \operatorname{tr} \Sigma_t
      = \frac{1}{n'} \operatorname{tr}\bigl(S'_t P_t S_t'^{\top}\bigr) + \sigma_t^2
      = \expect, e_t^2 ,
\end{equation}
where $e_t^2 = \frac{1}{n'} \sum_j \bigl(Y_j - \hat y_t(t_j; u)\bigr)^2$ and the
expectation is under the forecast law. When the actual measurements follow this law, the
score $e_t^2/V_t$ has mean one, and its whole distribution follows from the
eigenvalues $\lambda_k$ of $\Sigma_t$, since
$n' e_t^2 \overset{d}{=} \sum_k \lambda_k z_k^2$ with $z_k$ independent
standard normal. The gauge can be computed \emph{a posteriori} from actual patient records, we evaluate its relevance in Section~\ref{subsubsec:cohort}.

In view of developing the decision support tool, consider the lower tail of the forecast law. It gives the risk of hypotension of a target $y$ under the
control input $u_y$ (Section~\ref{subsec:command}), where the risk is the worst-case scenario of a single value under the threshold. This yields
\begin{equation}
  \label{eq:risk}
  \begin{aligned}
  R_t(y) &= \Pr\bigl(\min_j Y_j < 65\bigr) \\
  &= 1 - \expect \prod_j
  \Phi\Bigl(\frac{\hat y_t(t_j; u_y) + [S'_t\, \delta\coord]_j - 65}{\sigma_t}\Bigr),
  \end{aligned}
\end{equation}
the probability and the expectation being under the forecast law. Given
$\delta\coord$, the noise is integrated out exactly, and the expectation over
$\delta\coord$ is estimated by Monte Carlo (Appendix~\ref{app:risk}). Given a
level $\alpha$, the personalised target is the smallest target whose NE doses
keep the risk at most $\alpha$,
\begin{equation}
  \label{eq:target}
  y^\star_t = \min \bigl\{ y \in [65, \hat F(u_{\max})] : R_t(y) \le \alpha \bigr\},
\end{equation}
$\hat F(u_{\max})$ being the fit's MAP at equilibrium under the maximal rate.
It is found by bisection (Appendix~\ref{app:risk}). When even $\hat F(u_{\max})$ fails, no recommendation is given and the decision returns to the clinician. The target thus carries the fit's uncertainty into the dose: a well identified patient
gets a target close to the floor, a poorly identified one a higher target, or
none.

\begin{figure*}[t]
  \centering
  \includegraphics[width=\textwidth]{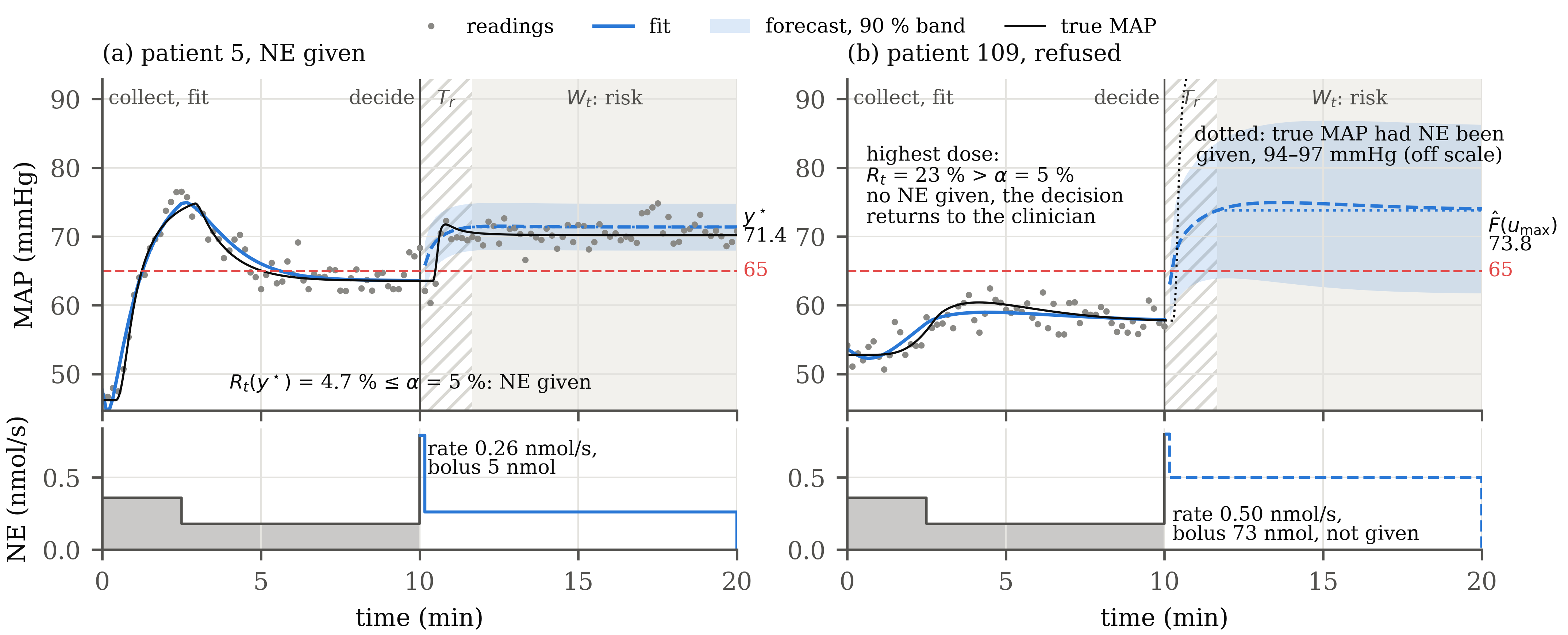}
  \caption{{One decision of the protocol at $t = 10$~min with
    $\alpha = 5\,\%$, on two virtual patients of Section~\ref{subsec:virtual}.
    Top: MAP readings (dots), the fit up to $t$ (solid blue), and from $t$
    the forecast of the readings under the dose evaluated, its mean (dashed)
    and pointwise 90\,\% band. Hatched, the rise $T_r$; shaded, the window
    $W_t$ on which the risk $R_t$ is computed; red, the 65~mmHg threshold.
    Bottom: NE rate, the collection common to all patients (gray) then the
    dose (blue; its bolus, given over 10~s, is drawn clipped and its amount
    written). (a) Patient given NE: the lowest target meeting the level,
    $y^\star = 71.4$~mmHg, has a risk of 4.7\,\%; its dose is applied to the
    true patient (black: true, noise-free MAP). (b) Patient refused: under
    the highest dose the fit allows, the forecast settles at 74~mmHg, but its
    band crosses 65~mmHg and $R_t = 23\,\% > \alpha$. No recommendation is given and the decision returns to the clinician. Dotted, the true MAP had the dose been given: the refusal comes from the fit's uncertainty, which ten minutes of
    data did not reduce enough.}}
  \label{fig:decision}
\end{figure*}

The two read-outs are checked where each can be. A real patient receives only
its recorded infusion, never the control input of $y^\star_t$, so its records check
the law itself through $e_t^2/V_t$ (Section~\ref{subsec:real}); the target is
checked on a virtual cohort, where control inputs can be applied and the true
patient is known (Section~\ref{subsec:virtual}).
{Fig.~\ref{fig:decision} shows both outcomes on two virtual patients.}

\subsection{Blood Pressure Control}
\label{subsec:command}

Given the fit $\est{\coord}_t$ and a target $y_{\mathrm{ref}}$ (in the protocol, each candidate target $y$ of~\eqref{eq:target}), we derive a control input that brings the MAP to $y_{\mathrm{ref}}$.
Classic control theory results, provided that the target $y_{\mathrm{ref}} \in C^2$, allow for exact tracking of a potentially time-varying target by model-inversion.
Because \eqref{eq:pd} is strictly increasing in $A_c$, each target $y_{\mathrm{ref}}$ corresponds to a unique central amount $A_c^{\mathrm{ref}} = h^{-1}(y_{\mathrm{ref}})$.
The peripheral amount then obeys the zero
dynamics $\dot A_p = k_{12}A_c^{\mathrm{ref}} - k_{21}A_p$, which are exponentially stable since
$k_{21} > 0$. Finally, \eqref{eq:pk} with $\beta = 1$ and $T_{\mathrm{lag}} = 0$
yields the depot amount and the input
\begin{equation}
  \label{eq:inversion}
  A_d = \frac{\dot A_c^{\mathrm{ref}} + (k_{12} + k_{el})A_c^{\mathrm{ref}}
        - k_{21}A_p}{k_a} , \qquad u = \dot A_d + k_a A_d .
\end{equation}

Elegant as this can be, manipulating a continuously varying infusion rate or a smooth $C^2$ target is not practical in a clinical setting. In practice, the NE injection is built from two primitives routinely used in surgery: a bolus, that is a high rate over a short period of time, and an infusion at a constant rate.
The constant infusion sets the equilibrium, and the bolus shortens the transient, to minimize the hypotensive time. The control input combines a constant component that sets the
desired equilibrium and an impulsive component that shapes the
transient response.
For a constant input $\bar u$, the equilibrium equations
$\dot{\state}=0$ yield $A_c^\star=\bar u/k_{el}$.
The input required to achieve the steady-state output
$y_{\mathrm{ref}}$ is therefore
\begin{equation}
	\label{eq:rate}
	\bar u = k_{el}\,h^{-1}(y_{\mathrm{ref}}).
\end{equation}

An ideal bolus of total dose $B$ corresponds to a Dirac impulse
$B\delta(s)$ superimposed on this constant input.
In practice, it is implemented as a rectangular pulse of duration
$\tau_b=10\,\mathrm{s}$ and amplitude $B/\tau_b$:
\begin{equation}
	\label{eq:cmd}
	u(t+s)=\bar u+\frac{B}{\tau_b}\,
	\mathbf{1}_{[0,\tau_b)}(s), \qquad s\geq 0.
\end{equation}
We seek the smallest admissible dose $B$ for which the first
hitting time of the target,
\[
\tau(B)=\inf\bigl\{s\geq 0:
h(A_c(t+s;B))\geq y_{\mathrm{ref}}\bigr\},
\]
satisfies $\tau(B)\leq T_r$.

Since $\tau(B)$ is nonincreasing in $B$ (Appendix~\ref{app:bolus}), the
minimum dose satisfying the hitting-time constraint is computed by bisection,
provided that the constraint is feasible within the admissible dose range.
Clinical maximums $u_{\max}$ and $B_{\max}$ ensure we give safe doses. They are set to the maximums used in the cohort of~\cite{joachimDesignPharmacokineticPharmacodynamic2024}, i.e. $0.5$~nmol/s and $75$~nmol, respectively.
{When even $B_{\max}$ misses $T_r$, the bolus is $B_{\max}$ and the MAP
reaches the target later.}

%% Experimental plan: real data (cohorts, results), virtual cohort (generation, results).
\section{Application}
\label{sec:application}
In this Section, we illustrate our approach on two datasets: the clinical cohort from~\cite{joachimDesignPharmacokineticPharmacodynamic2024} and a virtual cohort on which we can validate our risk-aware infusion protocol.
\subsection{Checking the forecast's law on the clinical cohort}
\label{subsec:real}

We first validate the forecast law of the readings ~\eqref{eq:readings},
on which our target~\eqref{eq:target} depends.
For this, we use the real cohort of~\cite{joachimDesignPharmacokineticPharmacodynamic2024}: we use the first part of each patient's data for fitting, and the remaining timeseries for validation.

\subsubsection{Cohort \& protocol description}
\label{subsubsec:cohort}
The cohort is that of~\cite{joachimDesignPharmacokineticPharmacodynamic2024}.
Each record holds the MAP, sampled every second for 17 patients and every 10~s
for 15, the NE given in boluses and infusions, and a few plasma concentration
samples, which only the fits of Fig.~\ref{fig:fits} use.
Every patient receives NE through a peripheral line, as one bolus (two for
three patients) of 35~nmol in median over 10 to 40~s, and an infusion whose
rate is changed a few times, 0.18~nmol/s in median and at most 0.6~nmol/s.
We split each record into windows of 2.5min.
At each decision time $t$, 5~min after the first reading and then every 2.5~min,
the reduced model is fitted to the MAP data up to $t$. This gives 155 windows from 31 patients.
Each window gives a forecast on $W_t$, which is scored by $e_t^2/V_t$, using the NE actually received during $W_t$.
The original cohort contains 10~s a,d 1~s records. The 1~s records are thinned to 10~s for comparison.

\subsubsection{Results}
For each clinical forecast window, we computed the predictive distribution
of the minimum MAP reading under the recorded NE input. If the forecast
is calibrated, the observed minimum should fall below its predicted
$\alpha$-quantile with probability $\alpha$. Across 155 windows from
31 patients, the observed frequencies were 4.5, 5.8, 8.4, and 15.5\,\%
for nominal levels of 1, 2, 5, and 10\,\%, respectively. In particular,
13 of the 155 observed minima fell below their predicted 5\,\% quantile.
These descriptive rates suggest that the forecast assigns too little
probability to low minimum MAP readings under the recorded inputs.
\label{subsubsec:valid-results}

\begin{table}[t]
  \caption{Score $e_t^2/V_t$ of the clinical cohort's windows, by decision
    time $t$ (min after the first reading): median with its 95\,\% interval
    (patients resampled), and 90th and 95th percentiles. A right law has a
    mean score of one.}
  \label{tab:real}
  \centering
  \scriptsize
  \begin{tabular}{lrccc}
    \toprule
    $t$ (min) & Windows & Median & 90\,\% & 95\,\% \\
    \midrule
    All          & 155 & 0.86 [0.54, 1.21] & 3.8 & 5.4 \\
    $[5, 10)$    & 60  & 0.46 [0.30, 0.79] & 1.4 & 1.9 \\
    $[10, 15)$   & 50  & 0.87 [0.55, 1.71] & 4.6 & 5.8 \\
    $\ge 15$     & 45  & 2.13 [1.13, 3.09] & 5.3 & 13.6 \\
    \bottomrule
  \end{tabular}
\end{table}

Table~\ref{tab:real} reports the median and upper percentiles of
$e_t^2/V_t$ by collection duration. Under the forecast law, this
score has expectation one. Its observed mean was 0.71 after
5--10~min of data collection and 3.42 after at least 15~min.
The forecast therefore tended to overestimate the squared error
in the earlier windows and to underestimate it in the later ones.
In the latter group, the median observed RMSE was 3.5~mmHg,
compared with a median predicted error scale $\sqrt{V_t}$ of
2.6~mmHg. We chose a 10-min collection for the virtual experiment
as a practical operating point, motivated by the intermediate-window
scores and the need to limit the time before the first recommendation.

\subsection{Full example on a virtual cohort}
\label{subsec:virtual}

\subsubsection{Virtual cohort generation and full protocol}
\label{subsubsec:generation}

% Rewritten 2026-11-07. Code: norcmd.cohort (draw, rules), norcmd.noise; driver
% opti.driver_target (--decisions 10): outputs/opti/target/virtual/
% all-p-reduced-3w-current-10min/summary.txt ("Draw:"); counts in numbers.md.
This section runs the whole protocol on a virtual cohort, where the `true' parameters of the
patients are known (but only used for comparison purposes). The input $u_{y^\star}$ corresponding to $y^\star$ defined by~\eqref{eq:target} is applied to each patient, and the complete model from\cite{joachimDesignPharmacokineticPharmacodynamic2024} is used to compute MAP values.
The patients are generated by drawing their twelve parameters independently
from the log-normal laws of~\cite{joachimDesignPharmacokineticPharmacodynamic2024},
$\theta_j = \bar\theta_j \exp(\omega_j z_j)$ with $z_j \sim \mathcal{N}(0, 1)$
(Appendix~\ref{app:coord}).
A draw is rejected only when it is implausible or leaves no room for control
(see Appendix~\ref{app:virtual} for details).
Draws from a fixed seed are kept until $N = 4{,}096$ patients are accepted,
enough to tell a 1\,\% rate of hypotension from 2\,\%.
The MAP is sampled every 10~s with a correlated noise of standard deviation
2~mmHg (Appendix~\ref{app:virtual}).

Every virtual patient receives the same collection: twice the median infusion
rate of the clinical cohort, 0.36~nmol/s, for the first 2.5~min, then the
median rate, 0.18~nmol/s until 10~min. As we discuss in Section~\ref{sec:discussion},
this design was settled empirically, and leaves room for a more informative, possibly personalized,
improvement.
At $t = 10$~min the reduced model is fitted to the simulated MAP values and $y^\star$ is computed from~\eqref{eq:target}
for $\alpha \in \{1, 2, 5, 10\}\,\%$. The corresponding control input~$u_{y^\star}$ is applied
from $t$, and the readings of $W_t$ give the events that $y^\star$ bounds.

Three other inputs are scored in the same way. The control (computed after the fit) corresponding to a fixed
70~mmHg is the protocol without the personalized target. The population input (computed from the model with $\coord = 0$) corresponding to a 70~mmHg target  gives the same dose to everyone. The oracle applies
rule~\eqref{eq:target} to the true patient, with the sensor's noise as its
only uncertainty: it gives the target of a perfect fit.

\subsubsection{Results}
In the virtual cohort, we applied the recommended input and counted
windows with at least one MAP reading below 65~mmHg. At a nominal risk
level of 5\,\%, the rule issued a recommendation for 74\,\% of the
patients. Among these patients, 6.9\,\% had a reading below 65~mmHg
(95\,\% interval: 6.0--7.8\,\%).

\label{subsubsec:virtual-results}

\begin{table}[t]
  \caption{Control inputs at $t = 10$~min on the 4{,}096 virtual patients: share of
    the patients for which the protocol is applied, share with a reading of $W_t$ below 65~mmHg with
    its 95\,\% interval (patients resampled), share whose true MAP,
    noise-free, goes below 65~mmHg, and median NE over the 10~min, nmol. Single
    target: the population's control input at the lowest target, the same for
    everyone, whose rate over the cohort is at most $\alpha$, found by trying
    every target on the whole cohort; none reaches 1 or 2\,\%. Fit at 70, $y^\star$'s patients: the
    fit's 70~mmHg control input on the patients $y^\star$ selects at 5\,\%.}.
  \label{tab:virtual}
  \centering
  \scriptsize
  \setlength{\tabcolsep}{3pt}
  \begin{tabular}{llcccc}
    \toprule
    Rule & $\alpha$ & \% selected & {\% reading $<65$} & {\% true $<65$} & {NE} \\
    \midrule
    $y^\star$ & 1\,\% & 67 & 2.3 [1.8, 2.9] & {0.04} & {111} \\
     & 2\,\% & 70 & 3.7 [3.0, 4.4] & {0.07} & {111} \\
     & 5\,\% & 74 & 6.9 [6.0, 7.8] & {0.13} & {112} \\
     & 10\,\% & 77 & 10.1 [9.0, 11.1] & {0.22} & {112} \\
    \midrule
    Oracle & 1\,\% & 93 & 0.9 [0.6, 1.2] & {0} & {100} \\
     & 2\,\% & 94 & 1.8 [1.4, 2.2] & {0} & {99} \\
     & 5\,\% & 95 & 4.6 [4.0, 5.3] & {0} & {96} \\
     & 10\,\% & 97 & 8.8 [7.9, 9.7] & {0} & {95} \\
    \midrule
    {Single, 89~mmHg} & {5\,\%} & {100} & {3.0 [2.4, 3.5]} & {1.3} & {313} \\
    {Single, 86~mmHg} & {10\,\%} & {100} & {10.0 [9.1, 10.9]} & {3.8} & {247} \\
    \midrule
    Fit at 70~mmHg & -- & 100 & 25.0 [23.8, 26.4] & {3.3} & {100} \\
    Fit at 70, $y^\star$'s & 5\,\% & 74 & 28.4 [26.8, 30.1] & 3.3 & 99 \\
    Population at 70 & -- & 100 & 65.7 [64.2, 67.2] & {51.5} & {57} \\
    \bottomrule
  \end{tabular}
\end{table}

{Table~\ref{tab:virtual} compares $y^\star$ with three rules that give NE
to every patient, from the easiest to beat to the hardest. The population's
dose for a 70~mmHg target leaves two thirds of the patients with a reading
below 65~mmHg. Fitting the patient first and aiming at the same 70~mmHg still
leaves a quarter of them. The hardest rule to beat aims at one higher target
for everyone, the lowest that keeps the cohort's rate below $\alpha = 5\,\%$:
89~mmHg. It meets the level, but it can only be found by trying every target
on the whole cohort, which the bedside never allows, and it needs nearly three
times the NE of $y^\star$. At the same level, $y^\star$ selects three quarters
of the patients, and 6.9\,\% of them have a reading below 65~mmHg.}

{But this might just be because $y^\star$ selects easy patients to give injections to.
To test for this hypothesis, we give the fit's 70~mmHg dose to the patients $y^\star$ selects.
A reading below 65~mmHg then occurs in 28\,\% of them, slightly more than the 25\,\% of the whole cohort. $y^\star$ don't just keep easy patients, its adapted target brings this share to 6.9\,\%.}

{Against the oracle, which applies~\eqref{eq:target} to the true patient,
$y^\star$ pays for the fit's uncertainty: it selects fewer patients and sets
higher targets. Even the oracle refuses a few patients, for whom no dose is
safe. And $y^\star$ delivers slightly less than it promises, with 6.9\,\% of
its patients below 65~mmHg against the 5\,\% asked, while the oracle stays
within its level. This gap points to the tails of the forecast law, already
too light on the clinical cohort (Section~\ref{subsec:real}), and is discussed
in Section~\ref{sec:discussion}. Most of these readings are the sensor's
noise: under $y^\star$, the true MAP falls below 65~mmHg for about one patient
in a thousand.}

\section{Discussion and Conclusion}
\label{sec:discussion}

\subsection{Discussion}
\subsubsection{Why the forecast is overconfident}

The forecast law is overconfident once the fit has seen enough readings. Indeed, on the
clinical cohort, the median score reaches 2.13 after 15~min
(Table~\ref{tab:real}). On the virtual cohort, $y^\star$ delivers 2.3, 3.7 and
6.9\,\% at $\alpha = 1$, 2 and 5\,\%, while the oracle, the same rule on the true
patient, keeps every level (Table~\ref{tab:virtual}). Four approximations of
Section~\ref{subsec:gauge} can cause it, which we now detail. 

\paragraph{The noise level.} $P_t$ treats $\sigma_t$ as known, though it is
estimated from the same readings, and a fit that follows part of the noise
underestimates it. On a run of the virtual cohort with a 20~min collection,
this explains about half of the excess: when $\sigma_t$ is still estimated but
never allowed below 2~mmHg, the true standard deviation of the virtual sensor,
$y^\star$ delivers 5.9\,\% at $\alpha = 5\,\%$ instead of 6.8\,\%, and stays
within its level at 10\,\%, at no measurable cost in control inputs.
% R082 arm B: sigma_t <- max(sigma_t, 2 mmHg) in the forecast's noise; R083: the
% same floor in P_t too adds nothing (5.84 against 5.86 %).

\paragraph{Linearization for uncertainty prediction.} The expansion~\eqref{eq:lin} ignores the curvature
of $\hat y$ and the truncation of the posterior by $\mathcal{B}$, and the
forecast carries it to a control input that the collection did not apply. We have not
measured its share. Sampling the posterior $\propto \exp(-J_t/2)$ on
$\mathcal{B}$ itself, for instance by importance sampling with the Gaussian
of~\eqref{eq:post} as proposal, would measure it on a subset of patients, at
the price of one simulation of the model per sample. This is a straightforward
extension of the protocol that we will explore in future work.

\paragraph{Model mismatch.}  In the clinical cohort, the predicted error falls
faster than the error measured on future MAP readings
(Table~\ref{tab:real}). Between 5--10~min and at least 15~min of data, the
median observed mean squared error $e_t^2$ falls from 25.8 to
12.1~mmHg$^2$, while the median predicted error $V_t$ falls from 52.2 to
6.7~mmHg$^2$. Most of that decrease
comes from reduced uncertainty in the fitted parameters and initial state,
whose median contribution falls from 41.1 to 1.0~mmHg$^2$. At that point,
$V_t$ is close to the estimated measurement-noise variance of
5.4~mmHg$^2$. The predictive covariance $\Sigma_t$ accounts only for
uncertainty in the fit and for measurement noise. A more precise fit can
still predict the wrong response to an infusion.
% numbers.md, the score's two sides (rerun of 2026-11-09).

When the model misses part of the patient's response, its error can persist
across successive MAP readings. We estimated an exponential correlation
model from the fit residuals and used it to whiten the output sensitivities
in the covariance calculation. This reduces the amount of independent
information attributed to closely spaced readings with similar errors.
Accounting for this correlation improved some screening decisions, but the
late forecasts remained overconfident.\footnote{The whitening comparison and the clinical correlation
diagnostic are in the code of this paper,
\url{https://github.com/SimonBure/risk-aware-bp-target}.}

In the virtual cohort, we generate the data with the complete model and
fit them with the reduced model. If the reduced model predicts a MAP that is
too high, it can underestimate the probability of a reading below
65~mmHg even after imposing the noise floor. By comparison, the virtual
sensor's own noise correlation has a small effect in a calculation with
constant predicted MAP and 50 readings at 10~s intervals. It changes the
crossing risk~\eqref{eq:risk} by less than 4\,\% relative at the four risk
levels considered, for example from 5 to 4.87\,\%.
% Sensor-only check: stationary Gaussian AR(1), rho = exp(-1), 50 readings.
% Set the common standardized threshold to Phi^{-1}(1-(1-alpha)^{1/50})
% and integrate the Gaussian transition density over the survival region.

Clinical validation data could help correct both the predicted MAP and its
uncertainty. If the patient's response changes over time, giving more
weight to recent measurements could also help the model track those changes.

\subsubsection{Collection informativeness}
Every virtual patient receives the same NE doses during the collection time, and
nothing guarantees that these are informative enough for a given patient, or for the dose
they will receive in the future. Even if designed empirically, the infusion plan (two rates in decreasing order) that we propose in
Section~\ref{subsec:virtual} proved better then an alternative designed after Joachim's
baseline (bolus followed by an infusion for 10~min). A search over over 165 candidates
(between one and four different infusions, administrated at random times) found none
clearly better than ours.

The challenge is more clinical than mathematical: even during the collection,
clinicians must keep the patient above 65~mmHg, and currently under ours, every
virtual patient spends some time below it. Evaluating and mitigating the effect of an early bolus will be investigated in future work. 

\subsection{Conclusion}
We proposed an online method that fits a reduced PK/PD model to MAP readings
and uses its forecast to select a personalised MAP target. In the clinical
cohort, the median forecast score was closest to one after 10--15~min and
exceeded one for longer records. At a nominal risk of 5\,\%, the method
recommended a dose for 74\,\% of the 4{,}096 virtual patients. Of these,
6.9\,\% had at least one MAP reading below 65~mmHg. With a fixed target of
70~mmHg applied to all virtual patients, the rate was 25\,\%.

Larger cohorts from routine intraoperative care could improve the population
model. These records are collected for clinical care, often without a protocol
designed to identify the response to NE. MAP can change after adjustments to
anaesthetic dosing, surgical stimulation such as pain, or an intravenous fluid bolus. These events can overlap with changes in NE delivery.

The question of collection informativeness also applies to these retrospective
data. We would exclude windows in which MAP varies too little relative to
measurement noise to identify the response dynamics, or in which the observed
changes are unlikely to be explained by NE. The selection should allow for
patient-specific delays and time constants. It should also account for the
fact that clinicians adjust NE in response to MAP, so the timing of a dose
change alone cannot establish the cause of a MAP change.

Synchronising MAP and NE delivery with other physiological signals, such as
heart rate and anaesthetic-depth measurements, and with time-stamped records
of drugs, fluids, and surgical events could help identify these disturbances.
We could then flag the affected windows or include documented interventions
as additional inputs to the model.

\section*{Acknowledgment}
Let us address a deep thank to Jona Joachim for providing its clinical dataset section
\ref{subsec:real} uses. We also thank him for his important clinical insights, that
helped orient our research towards what is most wanted by the clinicians.
These insights helped shape the paper better as well.

%% Appendices: the details of Sections II and III that the main argument does not
%% need. Placed after the acknowledgment, before the references, in the same
%% two-column format; they count toward the 6 pages.
\appendices

\section{Normalised coordinates and bounds}
\label{app:coord}

The normalised coordinates are $\eta_j = \ln(\theta_j/\bar\theta_j)/\omega_j$
for each parameter and $(\state_0 - \bar\state_0)/s_x$ for the three initial
amounts. Here $\ln\theta_j$ follows the Gaussian population distribution
of~\cite{joachimDesignPharmacokineticPharmacodynamic2024}: $\ln\bar\theta_j$ is
its mean and $\omega_j$ its inter-individual standard deviation. $\bar\state_0$ is
the state that the recorded infusion gives at $t_0$ under the population model, and $s_x$ is a
normalisation factor in nmol. Tthe population model hence gives $\coord = 0$, with its
prior $\mathcal{N}(0, I)$.

The box $\mathcal{B}$ of~\eqref{eq:fit} keeps each parameter within
$\pm 3\,\omega_j$ of $\bar\theta_j$, each amount between zero and a cap, and the
MAP without NE, $h(0)$, at least 30~mmHg, to avoid completely non-biological fits.

\section{Noise level $\sigma_t$}
\label{app:sigma}

The noise variance $\sigma_t^2$ is set by empirical
Bayes~\cite{mackay_bayesian_1992}: it maximises the marginal likelihood of the
readings under the expansion~\eqref{eq:lin}. Setting its derivative to zero
gives
\[
  \sigma_t^2 = \frac{\mathrm{SSE}_t}{n_t - p_t}, \qquad
  p_t = \operatorname{tr}(I - P_t),
\]
where $\mathrm{SSE}_t$ is the sum of squared residuals of the fit and
$p_t \le d$ the effective number of parameters that the readings determine.
Since $\mathrm{SSE}_t$ and $p_t$ depend on $\sigma_t$ through the fit, this
equation is solved by fixed-point iteration: starting from
$\sigma_t^2 = 3.99$~mmHg$^2$, the median mean squared residual of the reduced
model fitted to the whole records of the clinical cohort, each step
refits~\eqref{eq:fit} with the current $\sigma_t^2$ and replaces it by the
right-hand side, until it changes by less than 5\,\% (at most five steps). On
the clinical cohort, every fit converges in two or three steps. The caveat is that
the solution is a stationary point of the marginal likelihood, not necessarily its
maximum.

% Both sides depend on $\sigma_t$ through the fit, so the condition is iterated:
% from $\sigma_t^2 = 3.99$~mmHg$^2$, the median mean squared residual of the
% reduced model fitted to the whole records of the clinical cohort, \eqref{eq:fit}
% is refitted and $\sigma_t^2$ updated until it changes by less than 5\,\%, or
% for at most five iterations. A fixed point is a stationary point of the
% marginal likelihood, not necessarily its maximum.

\section{Computing the risk and the target}
\label{app:risk}

The expectation of~\eqref{eq:risk} over $\delta\coord = L_t \xi$, with
$P_t = L_t L_t^\top$ and $\xi \sim \mathcal{N}(0, I_d)$, is estimated with
$10^4$ draws of $\xi$, which leaves a Monte Carlo error of at most 0.2 point at
$\alpha = 5\,\%$. The target~\eqref{eq:target} is found by bisection to
0.05~mmHg, which assumes that the risk falls as the target rises: the forecast
mean rises with the NE dose, but $S'_t$ changes with it too.

\section{Monotonicity of the hitting time in the bolus}
\label{app:bolus}

The pulse~\eqref{eq:cmd} acts on the depot compartment, and its effect on $A_c$
is governed by the positive linear dynamics of the reduced
model~\eqref{eq:pk}. For a fixed initial state, $A_c(t+s;B)$ is therefore
nondecreasing in $B$ at every time $s$. Since $h$ is strictly increasing, the
same ordering holds for the MAP trajectories, and $\tau(B)$ is nonincreasing in
$B$. The smallest dose with $\tau(B) \le T_r$ can thus be found by bisection.

\section{Hypotensive Virtual patients}
\label{app:virtual}

A draw of the twelve parameters is rejected when:
\begin{itemize}
  \item NE lowers the MAP, $E_{\max} \le E_0$;
  \item the patient is not hypotensive without NE, or implausibly so:
    $F(0) \notin [30, 65]$~mmHg, $F(u) = h(u/k_{el})$ being the MAP at
    equilibrium under the rate $u$;
  \item 70~mmHg cannot be reached, $F(u_{\max}) < 70$~mmHg, so that a target
    above 65~mmHg remains within reach;
  \item the MAP exceeds 200~mmHg at equilibrium under $u_{\max}$, or under the
    heaviest input of the protocol: the maximal rate up to the decision, then
    the largest bolus on top of it.
\end{itemize}
The MAP noise is a stationary Ornstein--Uhlenbeck process of standard
deviation 2~mmHg and correlation $e^{-|\Delta t|/10\,\mathrm{s}}$, in line with
the residuals of the clinical cohort.

\bibliographystyle{IEEEtran}
\bibliography{refs}

\end{document}